# Modelling the effect of training on performance in road cycling: estimation of the Banister model parameters using field data

P. SCARF*[1], M. SHRAHILI[2], N. ALOTAIBI[3], S. JOBSON[4] and L. PASSFIELD[5]

[1]*Salford Business School, University of Salford, Salford M5 4WT, UK, email: p.a.scarf@salford.ac.uk*
[2]*Department of Statistics and Operations Research, College of Science, King Saud University, Riyadh, Kingdom of Saudi Arabia, email: msharahili@ksu.edu.sa*
[3]*Department of Mathematics and Statistics, Al Imam Mohammad Ibn Saud Islamic University, Kingdom of Saudi Arabia, nmaalotaibi@imamu.edu.sa*
[4]*Department of Sport and Exercise, University of Winchester, UK, email: simon.jobson@winchester.ac.uk*
[5]*Endurance Research Group, School of Sport and Exercise, University of Kent, UK, email: l.passfield@kent.ac.uk*
*\* corresponding author*

Abstract: We suppose that performance is a random variable whose expectation is related to training inputs, and we study four performance measures in a statistical model that relates performance to training. Our aim is to carry out a robust statistical analysis of the training-performance models that are used in proprietary software to plan training, and thereby put them on a firmer footing. The performance measures we consider are calculated using power output and heart rate data collected in the field by road cyclists. We find that parameter estimates in the training-performance models that we study differ across riders and across performance measures within riders. We conclude therefore that models and their estimates must be specific, both to the individual and to the quality (e.g. speed or endurance) that the individual seeks to train. While the parameter estimates we obtain may be useful for comparing given training programmes, we show that the underlying models themselves are not appropriate for the optimisation of a training schedule in advance of competition.

## 1. Introduction

Performance in sport is determined by tactics, equipment, technique, psychology, physical preparation, and chance (e.g. Moffatt et al., 2014; Bell et al., 2016; Kay, 2014; Shafizadeh and Gray, 2011; Baker and Cobley, 2008; Yee, 2014), and research has sought to quantify the effect upon performance of these factors. In many sports, physical preparation (which we shall call training) is the most important of these factors, and the effect of training upon performance is the most difficult to quantify (Jobson et al., 2009; Passfield et al., 2016). Mathematical models of the training-performance relationship exist (e.g. Avalos, 2003; Bull et al., 2000; Clarke and Skiba, 2013; Gouba et al., 2013; Bergstrom et al., 2014; Poole et al., 2016; Kolossa, 2017), but their application is limited because none are sufficiently individualised or well estimated. There is a very large literature on physiological adaptations to training (e.g. Hogan and Powers, 2017), but this also is not individualised. Therefore, state-of-the-art prescription of training is based on qualitative reasoning and coach and athlete experience. Thus, more, but not too much training is recognised as somehow optimal, and this notion was first formalised by Banister and co-workers (Banister et al., 1975; Calvert et al., 1976; Morton et al., 1990; Morton, 1997). Furthermore, much analysis of training relies on simply monitoring training input (e.g. power output and heart rate) and the response to training, whether performance output (e.g. time taken, distance thrown, power output developed) or physiological response (e.g. heart rate variation), using well-developed tools (e.g. power meters, heart rate monitors and associated software) (Plews et al., 2013; Halson, 2014; Gabbett et al., 2017; Metcalfe et al., 2017). However, the quantification of the effect of a unit of training on performance remains an open problem (Fister et al., 2015), and it is an important problem not just in sport (e.g. Csapo and Alegre, 2016).

There are many difficulties. The first of these is the quantification of training itself. Estimable measures of training, for example the training impulse or TRIMP (Banister and Calvert, 1980), are arguably too crude, and sophisticated measures are not estimable. The TRIMP measure defines the training load upon an athlete of a training session as the number of heart beats of the athlete in the session. Modified-TRIMP (Morton et al., 1990) puts more weight on instances of high heart rate, but has three parameters that an individual must estimate: two that specify what is a high heart rate and one that specifies the weighting scheme. The second difficulty is that the effects of individual training sessions are not additive (Alotaibi, 2017), although recent work has gone some way to addressing this (Kolossa et al, 2017). The third difficulty is the specification of a measure of performance. This may appear straightforward (e.g. time taken or distance thrown) but in many sports (e.g road cycling, rowing, soccer) performance other than simply winning is less tangible (Pyrka et al., 2011; Wimmer et al., 2011). The fourth difficulty is the quantification of the training-performance relationship must be individualised, because athlete A cannot expect the same performance response to a unit of training input as athlete B (Mujika et al., 1996; Avalos et al., 2003). And then finally, supposing that training and performance metrics are appropriately specified, the training-performance relationship is very noisy (Passfield et al., 2016).

The implication of this final point is that a performance metric must be frequently measured if it is to be useful. Therefore, in the approach that we describe in this paper, we analyse the power output and heart rate of cyclists collected regularly and routinely in the field using proprietary equipment. Thus the expectation is that a rider will collect data for every ride, training and competition, throughout their training history, and seek to use these data to monitor their performance and to plan training. We propose a number performance metrics and relate these to training using a statistical model first proposed in Shrahili (2014) and later refined in Alotaibi (2017). In so doing, we quantify the effect of a unit of training upon performance for each of a group of cyclists for whom data are available to us. The novelty of our methodology lies i) in the estimation of the training effect using routinely collected field data rather than using infrequent testing to measure performance, ii) in the necessary development of performance metrics that are appropriate for field data, and iii) in the quantification of the precision of model-parameter estimates. We compare the efficacy of our proposed performance metrics, and discuss how the training-performance model might be used to optimize training and the limitations of the methodology.

The structure of the paper is as follows. First we describe the field data. Section 3 discusses the quantification of training, and in section 4 we describe our proposed performance metrics. In section 5, we describe the training-performance model and its estimation. In section 6, we present our results and compare the performance metrics. We conclude with a discussion of practical implementation, limitations and potential development.

2. The data

The methodology we developed is illustrated using data of ten competitive, male road cyclists (Table 1). These riders collected data on power-output and heart rate for nominally all their sessions (training, testing and competition) over a period of between 6 and 14 months. Power output was measured using power-meter cranks (SRM, Julich, Germany); heart rate was measured using a heart rate monitor; the sampling interval for the two variables was 5 seconds (Figure 1). We refer to these data for each rider as the rider history. The data were obtained under standard ethics protocols. Average power for each session for each rider is shown in Figure 2. There was no information about missingness, and session data on a particular day is absent either if there was no rider or if the session was not recorded (due to e.g. change of bike or recording error).

Table 1: Rider characteristics (age, weight, height at start of data collection period)

| Rider | Age | Mass (kg) | Height (cm) | Mean power (W) | Power UQ (W) | Training period (days) | Training sessions |
|---|---|---|---|---|---|---|---|
| 1 | 45 | 74 | 183 | 203 | 291 | 160 | 112 |
| 2 | 52 | 75 | 175 | 203 | 307 | 249 | 88 |
| 3 | 35 | 71 | 181 | 214 | 291 | 287 | 108 |
| 4 | 42 | 78 | 179 | 201 | 246 | 317 | 112 |
| 5 | 21 | 61 | 171 | 197 | 280 | 273 | 101 |
| 6 | 27 | 72 | 184 | 265 | 384 | 338 | 146 |
| 7 | 40 | 76 | 178 | 222 | 323 | 272 | 152 |
| 8 | 34 | 77 | 182 | 165 | 274 | 348 | 162 |
| 9 | 34 | 88 | 186 | 167 | 214 | 333 | 197 |
| 10 | 29 | 72 | 175 | 187 | 260 | 410 | 251 |

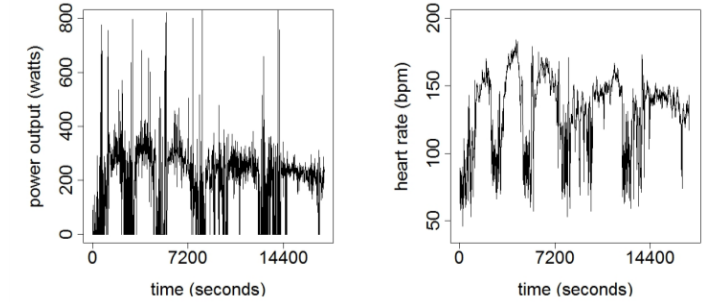

Figure 1. Example session: session 12, rider 3, ambient temperature 17.3 degC, duration 5 hours and 11 minutes; (left) power output vs time; (right) heart rate vs time.

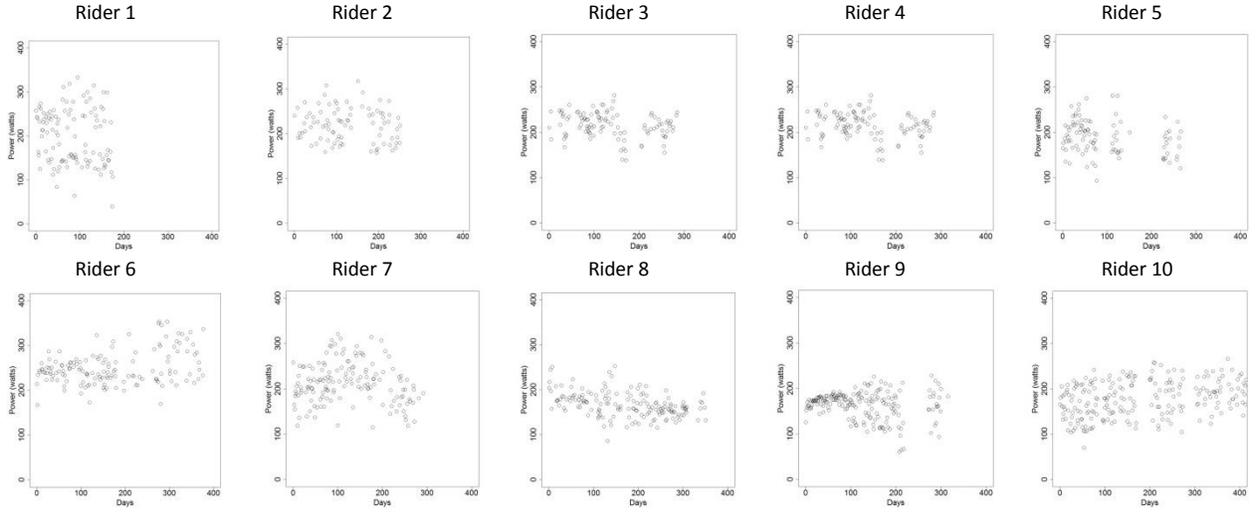

Figure 2. Average power output (watts) for each recorded session for each rider.

## 3. Quantification of training

Banister et al. (1975) and Calvert et al. (1976) proposed an additive model of training in which the preparedness or readiness to perform on day $i$ is the accumulation of the benefit (fitness) and detriment (fatigue) of preparatory training sessions:

$$W_i = W_0 + k_a \sum_{j=1}^{i-1} w_j e^{-r_a(i-j)} - k_f \sum_{j=1}^{i-1} w_j e^{-r_f(i-j)} . \qquad (1)$$

Here $W_i$ is the preparedness on day $i$; $w_j$ ($j = 1, \ldots, i-1$) are the known training loads of sessions on days prior to day $i$, both in the same arbitrary units; $W_0$ is the initial, baseline preparedness; $k_a < k_f$ are scale constants that control the size of the immediate training benefit with respect to the immediate training detriment; and $r_a < r_f$ are "discount rates" for the benefit and detriment of past training (although at the estimation stage we will parameterise the model with $\tau_a = 1/r_a$ and $\tau_f = 1/r_f$). For the training loads $w_j$ we use the training impulse (TRIMP) of the training session on day $j$, so that $w_j = d_j \bar{h}_j$, where $d_j$ and $\bar{h}_j$ are respectively the duration (in minutes) and the average heart rate (in beats per minute) of the training session. Note, no session on day $j$ corresponds to $w_j = 0$ in this notation. More generally, multiple sessions on day $j$ can be accommodated in the model, but not in this notation.

Without loss of generality, we can set $k_a = 1$ and $W_0 = 0$, so that response on day $t$ to a single training session on day 0 is

$$W_t = w(e^{-r_a t} - k_f e^{-r_f t}). \qquad (2)$$

In (2), initially (when $t = 0$), the response is $w(1 - k_f)$, so that the immediate, resultant effect of training (preparedness) is negative (since $1 = k_a < k_f$). Then the benefit and detriment decay, but the latter more quickly (since $r_a < r_f$), so that at time $t_0 = \log_e k_f / (r_f - r_a)$ the preparedness is zero (it has returned to baseline). Then at time $t^* = \log_e (k_f r_f / r_a) / (r_f - r_a)$ it is positive and maximum, and as $t$ increases it decays to the baseline (zero). The time interval between the preparedness being maximum and its decaying to half the maximum value (the pseudo-half-life) we denote as $t_{\text{half}}$ (but omit the expression for brevity). If $k_f = 2$ and $r_a = 1/8$ and $r_f = 1/2$ (days$^{-1}$) for example, then $t_0 = 1.8$, $t^* = 5.5$, and $t_{\text{half}} = 13.3$ (days). Thus, given the model and these parameter values, the optimum time to compete following a single bout of training is 5.5 days.

The above suggests that the Banister model (equation 1) can be used to optimise the timing of training inputs. However, suppose one performs two training sessions of equal training load (set to 1 unit); the first at time 0 and the second at time $s$. Then the response (preparedness) at time $t$ is

$$W_{t,s} = \left(e^{-r_a t} - k_f e^{-r_f t}\right) + \left(e^{-r_a(t-s)} - k_f e^{-r_f(t-s)}\right) = W_t + W_{t-s}.$$

Now $W_t$ is maximum at $t = t^*$, and so the maximum of $W_{t,s} = W_t + W_{t-s}$ must occur at $(t = t^*, s = 0)$ and its maximum value is $2W_t(t^*)$. Therefore, the model implies that training sessions should be performed concurrently. Practically, this is clearly neither possible nor sensible. The 6-parameter model proposed by Busso (2003), and studied by Kolossa et al. (2017), that allows for a non-linear response to training, has the same short-coming. The problem lies in the discretisation of the timing of training inputs, when in reality the cumulative effect of training on the fatigue induced by training is continuous (instantaneous). That is, the model (1) ignores the reality that the load that accumulates within a session affects the fatigue induced by the remainder of a session.

Nonetheless, the model is still useful for describing the relative sizes and durations of fitness and fatigue effects and for designing a training taper (the recovery period prior to competition). Therefore, researchers have sought to estimate its parameters, $k_f$, $\tau_a = 1/r_a$ and $\tau_f = 1/r_f$, by relating performance to training (preparedness) (Morton et al., 1990; Mujika et al., 1996; Busso et al., 1997, 2002; Millet et al., 2002; Hellard et al., 2006; Gouba et al., 2013). However, these attempts have not accounted adequately for sources of error in the training-performance relationship. We will develop a model to account for two sources of error. However, we first describe the measurement of performance.

## 4. The estimation of performance

We describe a number of performance metrics. Our purpose in doing so is to demonstrate that: i) although notionally one might suppose data allow the measurement of performance in fact they only allow the estimation of performance; and ii) the methodology we propose can use any performance measure that the data allow. On the first objective here, we suppose performance is measured with error, and we quantify this error in a statistical model. In this way, we distinguish between the notion of preparedness in the sports science literature (e.g. Busso and Thomas, 2006), which is the expectation of performance, and performance itself which is a random variable with this expectation.

This distinction between what an athlete "gives" during a session and what the same athlete might have "given" has been made by Shmanske (2011). Our first performance metric accounts for this gap in effort by estimating, for each session, the expected power output at a particular heart-rate threshold, denoted by $h_q$. We assume that the heart rate of an individual rider at time $t$ in session $i$, $H_{it}$, ($i = 1, ..., n$), is related to the power output $l$ time units earlier, $P_{i,t-l}$, by the equation

$$H_{it} = a_i + b_i P_{i,t-l} + cT_i t + \epsilon_t, \quad (3)$$

where $l$ is the heart rate lag, $a_i$ and $b_i$ are rider-session constants, $T_i$ is the ambient outside temperature for session $i$, $c$ is a global rider constant for cardiac drift, $n$ is the number of sessions in the rider's history, and $\epsilon_t \sim N(0, \tau^2)$ independently for all $t$. The literature justifies the linearity assumption (see e.g. Grazzi et al., 1999), the existence of the lag (see e.g. Stirling et al., 2008), for which we recommend the value $l$=15s (see Alotaibi, 2017), and the cardiac drift term, whereby, at a given power output, the heart rate drifts upwards at rate $cT_i$ (see e.g. Wingo et al., 2005). The model (3) is fitted by maximum likelihood to the rider history to obtain the estimates $\hat{a}_i$, $\hat{b}_i$ ($i = 1, ..., n$), and $\hat{c}$ and the covariance matrix estimate. Then, the performance metric for session $i$ is the estimated power output at the heart rate threshold $h_q$ (at a given reference time $t_R$ and a reference temperature $T_R$):

$$\hat{p}_{h_q,i} = (h_q - \hat{a}_i - \hat{c} T_R t_R)/\hat{b}_i. \quad (4)$$

The variance of $\hat{p}_{h_q,i}$ is estimated using the delta method. We set $t_R = 3600$s and temperature $T_R = 20°C$. As $h_q$ increases the variance of variance of $\hat{p}_{h_q,i}$ increases, so we recommend a *moderately* high value of $h_q$. We use the 75th percentile of heart rate of the rider, calculated over their entire history, denoted $h_{75}$. As a rider adapts to training, and all else being equal, we would expect $\hat{p}_{h_q,i}$ to increase. One might use this value of $h_q$ to specify a model for endurance training, and a higher value for sprint training, or use both in a multivariate performance metric.

The same model (3) can be inverted so that power output is a response to heart rate, and the estimated expected heart rate, $\hat{h}_{p_q,i}$, required to maintain a specified high power output $p_q$ in session $i$ is considered as the performance measure for session $i$. This measure is considered in detail in Alotaibi (2017), and briefly reported later in the results.

The next two performance measures take a somewhat different form, and use only the power outputs in a session. We estimate the "maximum" power maintained by a rider for $d$ seconds during the session. To do so, we first specify increasing power outputs $p_1, p_2, ...$ Then we determine from the data the longest duration $d_k$ for which the power output $p_k$ is sustained, and fit a parametric model to the pairs $(p_k, d_k)$ ($k = 1, 2, ..., m$). This model is

$$\log_e p_k = a_i + b_i \log_e d_k + \varepsilon_k,$$

where $\varepsilon_k \sim N(0, \tau^2)$ independently for all $k$. The rider-session constants $a_i$ and $b_i$ are estimated, $d$ is specified (e.g. $d = 10$s for sprint training, $d = 300$s for endurance training), and the corresponding "maximum" power is estimated: $\hat{p}_{d,i} = e^{\hat{a}_i} d^{\hat{b}_i}$. Figure 3 illustrates this procedure for a single session.

The final proposed performance measure is similar to the above and is based on the concept of the critical power model (see e.g. Clarke and Skiba, 2013) that relates power output to duration through

$$p = (p_0 - p_\infty) e^{-\theta d} + p_\infty + \text{error},$$

where $p$ is the power output that can be sustained for duration $d$, $p_0$ and $p_\infty$ are parameters, the former the "peak power" and the latter the critical power (the power output that can be notionally sustained indefinitely), and $\theta > 0$ so that the shorter the duration the greater the power output that can be sustained. The method then proceeds as with the maximum power metric, and assuming normal, independent errors the parameters are estimated using maximum likelihood. The performance metric for session $i$ is the estimated peak power, $\hat{p}_0$.

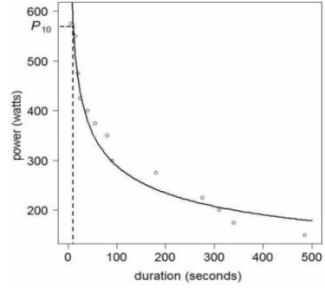

Figure 3. Observed power output against duration and fitted power-duration curve for a single training session for one rider

## 5. Estimation for the training-performance model

We consider two sources of error in our training-performance model. Firstly, given a particular trained state of the athlete (preparedness), the latent (unobserved) performance is a random variable with this preparedness as its expectation, but for arbitrary location and scale constants. Secondly, the measured (observed) performance is a random variable, because, for example, $\hat{p}_{h_q}$ in (4) is itself an estimate of the latent performance. For each source of error we assume normality for convenience. Thus for session $i$ ($i = 1, \ldots, n$) the latent performances are mutually independent and distributed as

$$P_i \sim N(\alpha + \beta W_i, \sigma^2). \tag{5}$$

We set $W_0 = 0$ in (1) since training up to day 1 is unquantified, so that training is comparative to the arbitrary baseline of zero on day 1. Further, for $i = 1, \ldots, n$ the observed performances are mutually independent and distributed as

$$\hat{P}_i \sim N(P_i, \lambda_i). \tag{6}$$

The distributional assumptions (5) and (6) imply that

$$\hat{P}_i \sim N(\alpha + \beta W_i, \sigma^2 + \lambda_i), \tag{7}$$

independently for all $i = 1, \ldots, n$. The variances $\lambda_i$ ($i = 1, \ldots, n$) are estimated using the delta method for the performance metrics $\hat{p}_{h_q,i}$ (equation 4), $\hat{h}_{p_q,i}$, and the maximum power $\hat{p}_{d,i}$, and obtained directly for the peak power performance measure, $\hat{p}_0$.

The parameters $\alpha$, $\beta$, $\sigma$, $k_f$, $\tau_a$ and $\tau_f$ are then estimated by maximising the log-likelihood function corresponding to (7):

$$\log L = -\frac{n}{2}\log(2\pi) - \frac{1}{2}\sum_{i=1}^{n}\log(\sigma^2 + \lambda_{s_i}) - \frac{1}{2}\sum_{i=1}^{n}\frac{\left(\hat{P}_{s_i} - \alpha - \beta W_{s_i}\right)^2}{\left(\sigma^2 + \lambda_{s_i}\right)}, \tag{8}$$

where $s_i$ is the day-number of session $i$, so that only days on which a session was recorded contribute to (8). Reasonable starting values for the parameters are needed in the numerical maximisation of (8). These were found by calculating the correlation between $\hat{P}_i$ and $W_i$ over a rough grid of values, and then using an interpolation technique to determine a point within the search space where the correlation was most positive. We also repeated the maximisation step from various starting points (not too far distant) to check for convergence to a global maximum.

## 6. Results and discussion

Table 2 presents the parameter estimates and their standard errors for all four performance measures. These estimates are comparable those obtained by others. For swimming, Mujika et al. (1996) reported $\tau_a$=41.4 and $\tau_f = 12.4$ (days) and $k_a$=0.062 and $k_f = 0.128$ (arbitrary units), and Hellard et al. (2006) reported $\tau_a = 38$, $\tau_f = 19$ (days), $k_a = 0.036$ and $k_f = 0.050$. Notice here that both $k_a$ and $k_f$ are estimated in spite of the fact that it is the relative sizes of these parameters that is important. Again for swimming, Gouba et al. (2013) reported $\tau_a = 42.3$ and $\tau_f = 15.3$ (days). For running, Morton et al. (1990) reported $\tau_a = 45$ and $\tau_f = 15$ (days) $k_a = 1$ and $k_f = 2$, accepting that $k_f$ should be twice $k_a$ (Banister et al., 1975), and Millet et al. (2002) reported $\tau_a = 20$ and $\tau_f = 10$ (days). For cycling, Busso et al. (1997) report $\tau_a = 60$, $\tau_f = 4$, $k_a = 0.0021$, $k_f$=0.0078 for participant A and $\tau_a = 60$, $\tau_f = 6$, $k_a = 0.0019$, $k_f$=0.0073 for participant B; Busso et al (2002) report values of $\tau_a$ from 30 to 60 days, and $\tau_f$ from 1 to 20 days. Kolossa et al. (2017) present values for 5 riders that show wide variability between individuals. None of these sources report standard errors.

Table 2. Parameter estimates (standard errors) for each of the four performance metrics.

| Rider | $\hat{p}_{h_{75}}$ | | | $\hat{h}_{p_{75}}$ | | | $\hat{p}_{10}$ | | | $\hat{p}_0$ | | |
|---|---|---|---|---|---|---|---|---|---|---|---|---|
| | $\hat{k}_f$ | $\hat{\tau}_a$ | $\hat{\tau}_f$ | $\hat{k}_f$ | $\hat{\tau}_a$ | $\hat{\tau}_f$ | $\hat{k}_f$ | $\hat{\tau}_a$ | $\hat{\tau}_f$ | $\hat{k}_f$ | $\hat{\tau}_a$ | $\hat{\tau}_f$ |
| 1 | 2.9 (0.5) | 35.3 (2.4) | 2.8 (0.9) | 1.7 (4.5) | 32 (17) | 1.6 (3.2) | 4.8 (1.3) | 33.6 (4.2) | 3.8 (2.1) | 1.1 (0.1) | 68.1 (12.6) | 5.2 (4.3) |
| 2 | 1.8 (0.5) | 31.2 (7.1) | 4.1 (0.8) | 4.0 (51) | 86 (65) | 0.1 (12) | 1.7 (0.3) | 105.8 (27.2) | 22.5 (1.7) | 1.3 (0.5) | 96.2 (69.7) | 11.6 (2.3) |
| 3 | 3.2 (0.4) | 16.3 (1.5) | 1.0 (0.1) | 4.7 (4.3) | 17 (6.1) | 1.4 (1.1) | 3.1 (0.6) | 67.9 (19.1) | 3.4 (2.3) | 1.8 (0.1) | 86.8 (1.5) | 2.2 (1.9) |
| 4 | 2.5 (0.2) | 95.1 (13.3) | 13.3 (0.9) | 1.1 (0.3) | 98 (27) | 33 (16) | 2.6 (2.5) | 57.9 (20.7) | 2.7 (2.4) | 4.7 (2.9) | 91.5 (29.9) | 12.2 (9.8) |
| 5 | 0.2 (0.1) | 140 (105) | 69.3 (3.5) | 1.4 (0.4) | 181 (109) | 125 (77) | 0.2 (0.1) | 83.9 (18.2) | 5.1 (1.6) | 0.7 (0.4) | 80.5 (44.8) | 4.8 (0.3) |
| 6 | 3.5 (0.1) | 162 (7.3) | 23.2 (0.4) | 2.0 (0.9) | 165 (51) | 27 (16) | 1.8 (0.3) | 87.9 (30.8) | 8.5 (1.9) | 1.2 (0.2) | 89.8 (25.8) | 3.4 (1.3) |
| 7 | 1.1 (0.2) | 132 (13.1) | 1.1 (0.5) | 7.1 (4.8) | 90 (62) | 0.1 (14) | 1.7 (0.5) | 76.3 (24.6) | 9.1 (5.9) | 3.1 (2.9) | 86.2 (25.6) | 11.1 (1.3) |
| 8 | 1.4 (0.2) | 162 (10.2) | 40.1 (2.3) | 1.1 (0.1) | 201 (79) | 161 (81) | 2.1 (0.4) | 84.7 (26.6) | 2.4 (0.6) | 3.2 (2.5) | 76.5 (34.9) | 2.2 (0.9) |
| 9 | 2.6 (1.8) | 169 (49.6) | 1.6 (0.2) | 3.2 (2.9) | 156 (78) | 8 (3.4) | 2.3 (1.2) | 100.5 (23.3) | 24.6 (11.5) | 2.4 (0.7) | 123.8 (23.9) | 19.8 (8.1) |
| 10 | 1.2 (0.1) | 126 (15.1) | 20.4 (1.5) | 2.0 (0.5) | 75 (22) | 34 (9.7) | 2.6 (0.4) | 90.8 (26.1) | 10.7 (1.5) | 1.1 (0.2) | 66.2 (58.4) | 4.9 (1.1) |

Thirdly, the very large values of the benefit decay constant imply that the training benefit is extremely persistent. All values of $k_f > 1$, implying that the immediate training detriment is larger than the immediate benefit, as required by the model.

The values of parameter estimates vary across the performance measures. Our view is that the first two measures, $h_{P75}$ and $P_{h75}$, better represent the variability in the training-performance than $p_{10}$ and $p_0$. We include the latter two because these have been extensively studied (e.g. Bull et al. 2000) and have more straightforward interpretations. However their estimation from the power-output data is ad hoc. A weakness of $h_{P75}$ is that it is rather counterintuitive, since as fitness increases, the performance measure decreases.

Figure 4 graphically illustrates the results for the $P_{h75}$ performance measure. Here we can observe variability between riders, variability in the training-performance relationship within riders (right hand plot) and to some extent how that varies with the number of sessions for which data are available, and the progression of preparedness (the solid line in the left-hand plot for each rider). Indeed there is large within and between rider variability, indicating the importance of presenting results about the precision of estimates. We can conclude from this also that parameter estimation should be individualised.

Over the training period for which data were collected, the difference in heart rate between when a rider was most trained (maximum *W*) and least trained (minimum *W*) is between 6% and 23% (excluding rider 5 because of the lack of his data). These are similar to values reported by others (Foster et al., 1996; Gabbett & Domrow, 2007), where a ten-fold increase in training load is associated with 10% improvement in performance. The progression in preparedness is summarised in Table 3 for each rider and for each performance measure. The quantity $\hat{\beta} \times \Delta W_{\max}$ is the change in preparedness (expected performance) between when the rider is least trained and most trained during the period of data collection. This number to some extent serves as a measure of the practical significance of the model, with a large value indicating a useful response to training. Riders show variable progression, and progression as a whole varies between the measures. These results reinforce the notions that estimation must be individualised and a performance measure should be carefully chosen.

## 7. Conclusion

In this paper, we study the role of four performance measures in a statistical model that relates performance output to training input. In particular, we suppose that performance is a random variable whose expectation is related to training inputs. This is a notion that has not been acknowledged in the literature to date. We use the term preparedness for the expectation of performance. We argue that it is important to distinguish these two concepts.

Our study is also important because training-performance models are used in proprietary software to plan training. It is our view that these models require a firmer statistical footing, through a robust analysis of them, and a quantification of the precision of estimation. This paper attempts such an analysis.

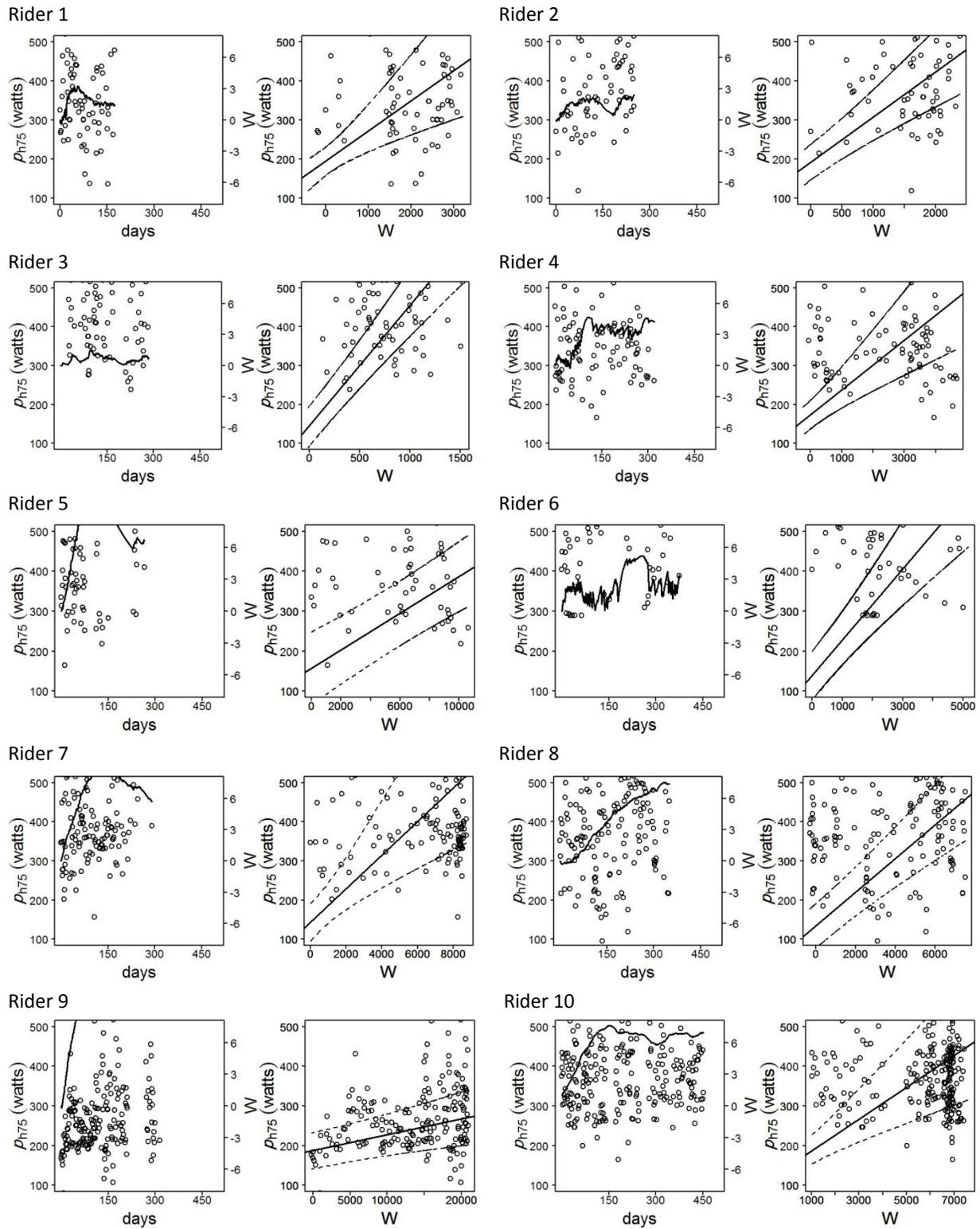

Figure 4. Two plots for each rider: left plot $\hat{p}_{h_{75}}$ (symbols) vs time in days and estimated preparedness $\hat{W}$ (line) vs time in days; right plot $\hat{p}_{h_{75}}$ vs estimated preparedness $\hat{W}$ (all sessions) with fitted line and 95% confidence bands for fitted line.

Table 3. Maximum preparedness improvement, $\hat{\beta} \times \Delta W_{\max}$, for each performance measure

| Rider | $\hat{p}_{h_{75}}$ | $\hat{h}_{p_{75}}$ | $\hat{p}_{10}$ | $\hat{p}_0$ |
|---|---|---|---|---|
| 1 | 29.8 | -10.6 | 47.4 | 9.8 |
| 2 | 10.9 | -21.7 | 124.3 | 25.1 |
| 3 | 7.3 | -12.0 | 36.4 | 11.4 |
| 4 | 15.5 | -20.7 | 36.3 | 12.6 |
| 5 | 19.0 | -3.4 | 63.8 | 35.9 |
| 6 | 30.7 | -23.0 | 167.6 | 17.5 |
| 7 | 12.4 | -10.8 | 117.6 | 3.7 |
| 8 | 5.3 | -9.2 | 78.0 | 6.4 |
| 9 | 12.8 | -7.1 | 129.6 | 6.1 |
| 10 | 15.3 | -8.2 | 88.8 | 1.6 |

The performance measures we consider are calculated using power-output and heart-rate data collected in the field. These measures depend on specific performance concepts that are different from one to another. These measures are related to a given training input measure. In each case, the parameters of the training-performance model are estimated, along with their standard errors. In the literature published to date, there has been a failure to acknowledge the statistical variability of estimates. Indeed, estimates have been presented in a way that overestimates their precision. We model two sources of imprecision, one whereby performance itself is variable (performance is a random variable whose expectation is preparedness), and the second whereby performance is measured with error.

Cycling lends itself to the model development and analysis we describe because power-output and heart-rate measurements can be routinely collected. Routine measurement of the latter is practical for most sports in both competition and training. Routine measurement of the former is not practical for most sports in both competition and training, either because performance (other than simply winning) is not well defined or because performance can be measured only infrequently.

Broadly speaking, we find that estimates across riders are dissimilar, while estimates across the performance measures but within riders are similar. On this basis, we conclude that training-performance models should be individualised and that general estimates should not be used to plan training. Choice of the performance measure is also a matter of individualisation—it really depends whether the athlete is seeking to improve speed or endurance. Also, we feel that the performance measures that use both power-output and heart-rate better model the nature of training. With these measures, the speed or endurance question reduces to one of choice of percentile 50, 75 or 90. However, choice of a very high threshold will lead to some difficulties with estimation as the training data available at high thresholds may be sparse.

At the outset of this study, existing training-performance models were presumed useful for planning training and that once parameters were estimated, training schedules could be optimised. This has turned out to be not the case. The discretisation of the timing of training inputs in these models implies that multiple bouts of training should be carried out simultaneously in order to maximise preparedness. The models are therefore inadequate, and a continuous time model is necessary to model training load accumulation within a session.

We focus on four different performance measures. Other choices are possible. Other time lags in the heart rate response to the power output might be studied. Estimation of parameters of the training input measure is another issue that requires further research, not least because in existing studies the imprecision of estimates is ignored. The impact of "missingness", training sessions for which data are missing, is another important issue for future study.

A*cknowledgements*: The authors would like to thank the Deanship of Scientific Research at King Saud University for funding this research under research contract number RG-1438-086. The project was also supported in part by EPSRC grant number EP/F006136/1.